\documentclass{article}
\usepackage[preprint]{neurips_2024}

\usepackage[utf8]{inputenc}
\usepackage[T1]{fontenc}
\usepackage{hyperref}
\usepackage{url}
\usepackage{booktabs}
\usepackage{amsfonts}
\usepackage{nicefrac}
\usepackage{microtype}
\usepackage{xcolor}
\usepackage{graphicx}
\usepackage{amsmath}

\usepackage[ruled]{algorithm2e}

\newcommand{\eps}{\varepsilon}

\newcommand{\set}[2]{\left\{#1\,\left\vert\; #2\right.\right\}}
\newcommand{\E}{\mathrm{E}}
\newcommand{\Var}{\mathrm{Var}}
\newcommand{\Cov}{\mathrm{Cov}}

\newtheorem{theorem}{Theorem}
\newtheorem{lemma}{Lemma}
\newtheorem{corollary}{Corollary}

\title{Reach Measurement, Optimization and Frequency Capping In\\Targeted Online Advertising Under k-Anonymity}

\author{%
  Yuan Gao\\
  LinkedIn Corporation\\
  Sunnyvale, CA, USA\\
  \texttt{yugao@linkedin.com}
  \And
  Mu Qiao\\
  LinkedIn Corporation\\
  Sunnyvale, CA, USA\\
  \texttt{mqiao@linkedin.com}
}

\begin{document}

\maketitle

\begin{abstract}
The growth in the use of online advertising to foster brand awareness over recent years is largely attributable to the ubiquity of social media. One pivotal technology contributing to the success of online brand advertising is frequency capping, a mechanism that enables marketers to control the number of times an ad is shown to a specific user. However, the very foundation of this technology is being scrutinized as the industry gravitates towards advertising solutions that prioritize user privacy. This paper delves into the issue of reach measurement and optimization within the context of $k$-anonymity, a privacy-preserving model gaining traction across major online advertising platforms. We outline how to report reach within this new privacy landscape and demonstrate how probabilistic discounting, a probabilistic adaptation of traditional frequency capping, can be employed to optimize campaign performance. Experiments are performed to assess the trade-off between user privacy and the efficacy of online brand advertising. Notably, we discern a significant dip in performance as long as privacy is introduced, yet this comes with a limited additional cost for advertising platforms to offer their users more privacy.
\end{abstract}

\section{Introduction}
\label{section:introduction}

Digital advertising now represents a significant proportion of the global ad expenditure \citep{StatistaAdSpend}. Thanks to its innate capacity for targeted and personalized ad delivery, digital advertising has become an effective platform for performance-based marketing. The key to effectiveness lies largely in relevancy. Initially, marketers identify potential customers by delineating their audience according to user geolocation and interests. This process is further refined as digital advertising platforms match users and ads based on real-time relevance, preventing advertisers' budgets from being squandered on less-engaged audiences. Furthermore, major online publishers now offer automated tools to aid advertisers in optimizing for their desired outcomes, such as purchases or app downloads. The capacity to measure and drive immediate impact via automated systems forms the backbone of the swift growth of online advertising.

Conversely, traditional marketing places substantial emphasis on brand advertising. The objective of brand advertising, unlike performance-based marketing, is to nurture long-term relationships with customers. This aim is pursued through consistent ad exposure to audiences who demonstrate potential interest in the brand. However, unlike performance-based advertising where more conversions translate to better outcomes, increased impression volume does not always yield desirable results in brand ads. Studies indicate that ad over-exposure negatively impacts the brand \citep{broussard2000advertising, ma2016user}. In response to this issue, the industry often employs frequency capping. Specifically, in a brand advertising campaign, a frequency cap of $c$ ensures an ad from the campaign is shown a maximum of $c$ times to the same user. In brand marketing, it's typical to define the campaign objective as maximizing impressions, subject to a frequency cap constraint. The objective is commonly referred to as reach when $c = 1$, as it represents the number of unique targeted users served. However, for simplicity, we refer to the general objective as reach in this paper, reserving the term unique reach for the case when $c = 1$.

A key competitive advantage of online advertising platforms over traditional media is their ability to enforce frequency capping at the individual user level. Yet, as privacy concerns escalate, tracking individual users may no longer be a viable option in the future \citep{iwanczak2022future}. Privacy protecting protocols are being implemented by leading online advertising platforms to protect the privacy of their user data \citep{linkedinprivacy, googleprivacy}. These measures leverage generalization techniques to ensure $k$-anonymity \citep{kanonymity}, a feature that makes an individual user indistinguishable from a group of other $k-1$ users. This paper delves into the reach measurement and optimization issue within this privacy paradigm.

\subsection{Related Work}
Traditional reach optimization research often operates under the assumption of known user identities, with a focus on challenges such as frequency capping and constrained ad matching problems \citep{mehta2007adwords, feldman2009online, buchbinder2014frequency, hojjat2017matching}. More practically, modern online advertising platforms often run under popular auction formats such as first and second price auctions, and campaign optimization under these auction formats are usually studied under optimal bidding \citep{balseiro2019learning, gao2022bidding, maehara2018optimal, zhang2014optimal}. However, these methodologies are met with significant hurdles when applied in scenarios where privacy protection protocols obscure full knowledge of user identities.

The preservation of privacy in digital advertising has attracted considerable research interest, with a significant focus on the issue of cross-publisher reach measurement in environments lacking third-party cookies. Influential studies propose the establishment of common identifiers across publishers used exclusively for aggregate measurement \citep{crossmedia2020}. Notably, \citep{skvortsov2019virtual} suggests a method for assigning virtual identifiers across various domains. With these unified identifiers in place, it becomes possible to amalgamate measurement statistics from different publishers using differentially-private solutions, which rely on either global \citep{kreuter2020} or local \citep{Wright2021} summarization algorithms.

Additional research avenues include exploring the intermediary stage where partial cross-publisher activity data is available from users who have given their consent to be tracked \citep{sang2022feedback}. In such instances, machine learning algorithms have been proposed as a means of learning from this data to infer and optimize non-trackable requests.

Distinguished from the existing body of work, this paper anticipates future scenarios where user privacy is stringently maintained, even within the context of first-party domains. The aim of this work is to evaluate the efficacy of measurement and optimization strategies under a specific privacy protocol, namely the $k$-anonymous privacy protocol.

\subsection{The $k$-anonymous Privacy Protocol}

\begin{figure}[t]
\centering
\includegraphics[trim=0 125 125 0, clip, width=0.95\linewidth]{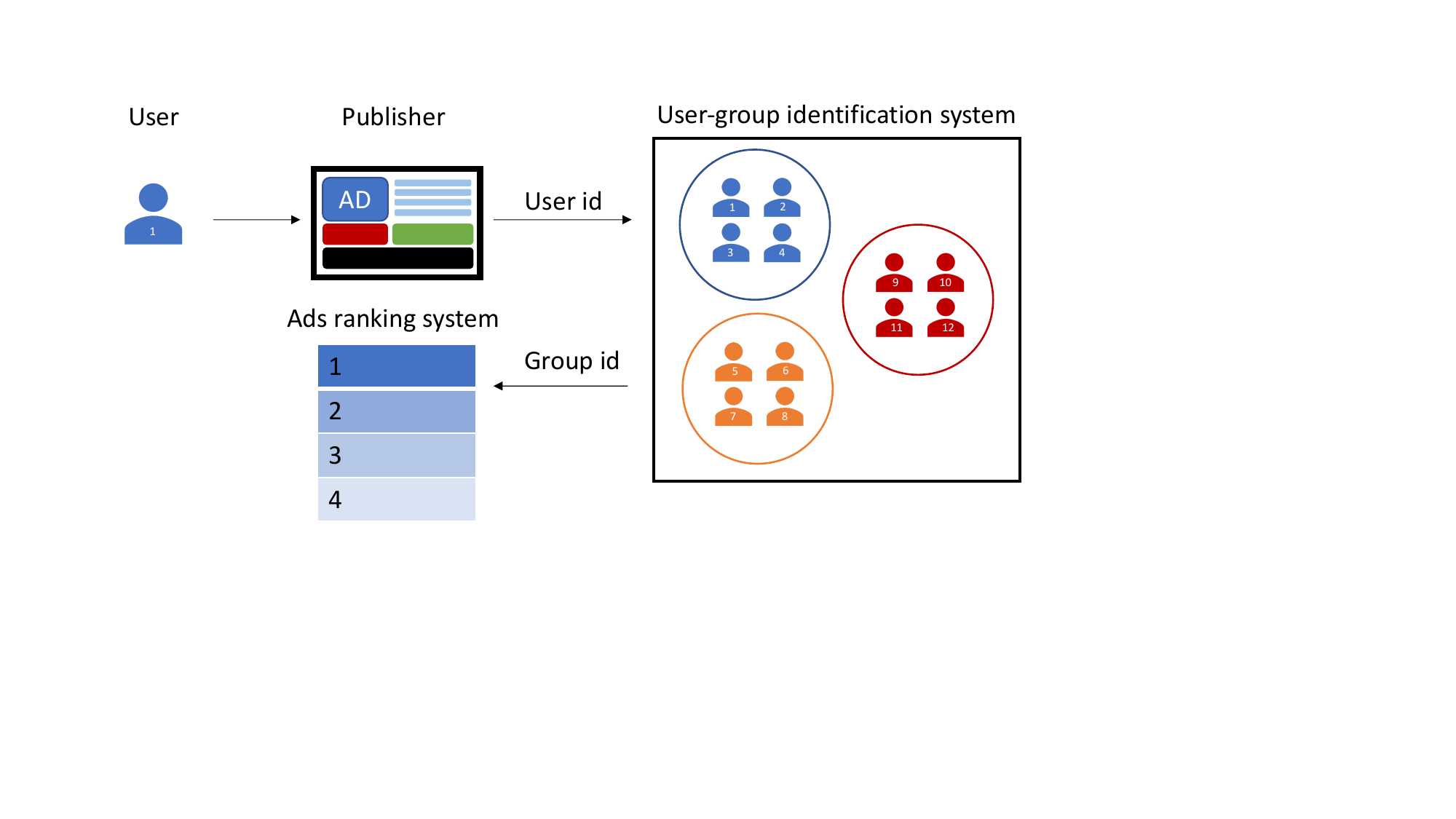}
\caption{Illustration of the k-anonymous Privacy Protocol, highlighting the process from user visit to ad display within the framework of $k$-anonymity.}
\label{fig0}
\end{figure}

In the realm of online advertising, the concept of $k$-anonymity plays a pivotal role in balancing the effectiveness of targeted advertising with the imperative need for user privacy. This protocol is adeptly implemented by publisher platforms, such as social media sites, search engines, or content providers, which meticulously form user groups \citep{epasto2021clustering}. Each group comprises $k$ members, safeguarding individual user actions by making them indistinguishable from $k-1$ others within the same cohort. These groups are thoughtfully created based on common user attributes like geolocation and interests. When a user visits the publisher and triggers an ad request, their information is forwarded to a separate identification system, which returns one of the groups containing the user. This group identity is then utilized by the downstream advertising system to make decisions on ad displays, ensuring that only group-level information is used, in order to maintain user anonymity (Figure \ref{fig0}).

While the optimal computation of $k$-anonymity is recognized as an NP-hard challenge\citep{MeyersonWilliams2004}, pragmatic algorithms have emerged \citep{DePascale2023, Bonizzoni2013}, facilitating its application in diverse real-world settings. Furthermore, the concept of $k$-anonymity seamlessly extends to certain advertising contexts, such as television ads targeted at households. Here, the ad is directed at a family unit, inherently forming a $k$-anonymous group, where the television network, by default, ensures a level of $k$-anonymity by targeting the household collectively rather than its individual members.

It's important to delineate the distinct roles of the publisher platform and its advertising system within this privacy-centric framework. The publisher platform is entrusted with the development and management of the $k$-anonymity algorithm and the formation of user groups, underscoring its commitment to privacy. Conversely, the advertising system operates under the privacy constraints established by these groups, aiming to deliver personalized advertising content in a manner that respects user anonymity.

From the vantage point of the advertising system, our objective is to navigate the intricacies of $k$-anonymity to optimize ad delivery while upholding the privacy standards set forth by the publisher platform. This involves devising strategies that maximize advertising effectiveness within the bounds of $k$-anonymous groups, ensuring that ads are relevant and impactful, yet privacy-compliant. By embracing this challenge, we contribute to the sustainable evolution of online advertising, where the personalization will coexist with privacy.

\section{Notations}

We use $i$ to index individual users, $j$ to indicate user groups, and $t$ to signify ad requests originating from a user group. As the primary focus of this research is the measurement and optimization of reach for a single ad campaign, there's no need to introduce additional indices for multiple campaigns. For the campaign in question, we assume its frequency cap is $c$ and it targets a total of $N$ users. For simplicity, we presume that the targeted user indices range from $1$ to $N$. These targeted users are assigned to $M$ distinct user groups, which collectively generate a total of $T$ ad requests.

Each ad request $t$ is generated by an unidentified user. However, the system has access to $j_t$, the index of the user group generating ad request $t$. We use $p_t$ to denote the probability that ad request $t$ comes from a user who has previously been served fewer than $c$ times.

We define $U_j$ as the set of targeted user indices in group $j$. Formally,
\[
U_j := \set{i}{
\begin{array}{c}
\texttt{user $i$ is in user group $j$} \\ 
\texttt{$\land$ user $i$ is targeted}
\end{array}
},
\]
We also define $G_i$ as the set of group indices containing user $i$:
\[
G_i := \set{j}{\texttt{user $i$ is in user group $j$}}.
\]
We use the random variable $X_i$ to denote the number of times user $i$ has viewed the ad.

For any given set $S$, we denote its cardinality as $|S|$. We represent the probability density function (p.d.f.) and cumulative density function (c.d.f.) of a Binomial distribution $\mathcal B(n, p)$ using $f(\cdot; n, p)$ and $F(\cdot; n, p)$ respectively, i.e., for all $x \le n$,
\[
f(x; n, p) := {n\choose x}p^x(1-p)^{n-x},
\]
and
\[
F(x; n, p) := \sum_{i=0}^x {n\choose i}p^i(1-p)^{n-i}.
\]
Here we extend the domain of $f$ and $F$ to $x > n$ by defining
\[
f(x; n, p) := 0, \ \texttt{and}\  F(x; n, p) :=  1, \ \forall x > n.
\]
Finally, $\phi$ and $\Phi$ denote the p.d.f. and c.d.f. for the standard normal distribution, respectively.

\section{Reach Measurement}
\label{section:measurement}

Under $k$-anonymity, the ability to track users at an individual level is compromised, making accurate reach measurement challenging. In this section, we focus on estimating reach when only the user group impressions are accessible. It's essential to keep in mind that these impressions result from auction-based bidding. The specifics of the reach optimization problem are outlined in Section \ref{section:optimization}.

As discussed in the introduction, typically, an ad campaign operates under a pre-determined frequency cap $c$, where reach is defined as the total number of impressions served within this cap. However, advertisers often require the reach metric for a variety of frequency caps. For instance, an advertiser may set the frequency cap at $c = 2$ to ensure that targeted users don't see the ad more than twice. Still, it would be beneficial for the publisher to report the unique number of users served by the ad.

Time is another important factor. Advertisers frequently assess campaign reach during the campaign flight, necessitating real-time updates of reach metrics. Additionally, reach metrics should accommodate range-based queries to enable advertisers to access reports for any given time frame, such as reach within the past week or month.

Formally, a time window identifies a subset of indices from the set ${1, \ldots, T}$. Let $n_j$ be the number of impressions for the ad campaign from user group $j$ within this window. The theorem below details the expected reach under a given frequency cap.

\begin{theorem}[Reach in Expectation]\label{thm:measurement}
The expected reach under a given frequency cap $c \ge 1$ is given by
\begin{equation}\label{eq:measurement}
\sum_{i=1}^N \left(\sum_{l=1}^{c} lf\left(l;\sum_{j \in G_i}n_j, \frac{1}{k}\right) + c - cF\left(c;\sum_{j \in G_i}n_j, \frac{1}{k}\right)\right).
\end{equation}
\end{theorem}

In the special case where $c = 1$, the following Corollary applies:
\begin{corollary}\label{cor:measurement_unique_reach}
Let $n_j$ be the number of impressions of the ad campaign from user group $j$, then the expected number of unique reach is
\begin{equation}\label{eq:unique_reach_measurement}
\sum_{i=1}^N\left(1 - \left(1 - \frac{1}{k}\right)^{\sum_{j \in G_i}n_j}\right).
\end{equation}
\end{corollary}
For the other extreme, as $c \rightarrow \infty$, equation \eqref{eq:measurement} simply becomes $\frac{1}{k}\sum_i\sum_{j \in G_i}n_j$, the expected total impressions. This number is bounded above by $\sum_j n_j$, the total group impressions. This upper limit is reached when all users within the groups are targeted.

It's also worth noting that in the degenerate case where $k = 1$ and each user belongs to a distinct group, Theorem \ref{thm:measurement} and Corollary \ref{cor:measurement_unique_reach} provide deterministic reach counts.

Theorem \ref{thm:measurement} allows the publisher to compute reach metrics for any given time window. In addition, the following alternative formulation of \eqref{eq:measurement} also facilitates a streaming algorithm that updates the expected reach as new impressions are observed, with a time complexity of $O(kc^{2})$ and a space complexity is $O(cT)$.
\begin{corollary}\label{cor:alternative_form}
The following is an alternative form of the expected number of reach under a frequency cap $c$. In other words, \eqref{eq:measurement} can be written as
\[
\sum_{i=1}^N \left(c - \sum_{l=0}^{c - 1} (c - l)f\left(l;\sum_{j \in G_i}n_j, \frac{1}{k}\right)\right).
\]
\end{corollary}

\subsection{Bounds on the Estimate}
Apart from consistently reporting the correct reach metric in expectation, it is also crucial to understand the extent to which the actual reach can diverge from the expectation. For instance, when the groups do not overlap, the reach's lower bound $R(c)$ is simply $\sum_{j = 1}^M \min(n_j, c)$, achievable when requests from each group originate consistently from the same user. Conversely, the upper bound of $R(c)$ is $\sum_{j = 1}^M \min(n_j, c|U_j|)$, realizable when, for example, targeted users in a group are catered to in a round-robin manner. The upper and lower bounds for the general case involving overlap can be complex to represent mathematically, but they can be computed using greedy algorithms.

When it comes to concentration bounds on the reach metric $R(c)$, it is significant to acknowledge that although $R(c) = \sum_i R_i(c)$ is a sum of random variables, Hoeffding's inequality does not apply in this context since $R_i(c)$ are typically negatively correlated for users within the same group. Instead, we can calculate the variance of $R(c)$ and apply Chebyshev's inequality to derive probabilistic bounds on the deviation. However, the analytical form of the variance for general $c$ is of combinatorial nature. Consequently, we present an explicit bound on $R(1)$. It hinges on the following lemma. For notation simplicity, we denote $R := R(1)$ as the reach when $c = 1$.

\begin{lemma}[Covariance]\label{lem:cov}
Let $R_i = \min(X_i, 1)$, where $X_i$ represents the number of impressions from user $i$. For two users $i, i'$, let $G_\cap := G_i \cap G_{i'}$ represent their intersecting groups and $G_\cup := G_i \cup G_{i'}$ their union. If $n_j$ denotes the number of impressions of the ad campaign from user group $j$, then the covariance $\Cov[R_i, R_{i'}]$ is given by
\begin{equation}
\begin{aligned}
\left(\left(1 - \frac{2}{k}\right)^{\sum_{j \in G_\cap}n_j} - \left(1 - \frac{1}{k}\right)^{2\sum_{j \in G_\cap}n_j}\right) \times \\
\left(1 - \frac{1}{k}\right)^{\sum_{j \in G_\cup - G_\cap}n_j}.
\end{aligned}
\label{eq:cov}
\end{equation}
\end{lemma}

Together with the observation that in this scenario $\Var[R_i] = \E[R_i](1 - \E[R_i])$, which equates to
\begin{equation}\label{eq:var}
\left(1 - \left(1 - \frac{1}{k}\right)^{\sum_{j \in G_i}n_j}\right)\left(1 - \frac{1}{k}\right)^{\sum_{j \in G_i}n_j},
\end{equation}
we obtain the following theorem.

\begin{theorem}[Concentration Bounds]\label{thm:concentration}
The probability that $R$ deviates its expectation $\E[R]$ (given in Theorem \ref{thm:measurement}) by $\epsilon\sigma$ is bounded by
\[
P\left(|R - \E[R]| \ge \epsilon\sigma\right) \le \frac{1}{\epsilon^2},
\]
where
\begin{equation}
\sigma^2 = \sum_i \Var[R_i] + \sum_{i \neq i'}\Cov[R_i, R_{i'}],
\end{equation}
where $\Var[R_i]$ is given by \eqref{eq:var} and $\Cov[R_i, R_{i'}]$ is given by \eqref{eq:cov}.
\end{theorem}

For general $c > 1$, we can compute the variance of $R(c)$ using Monte Carlo methods. For more information, refer to Section \ref{section:experiments}. Finally, the subsequent theorem offers estimates on over-exposure, crucial for advertisers concerned about its negative impact.

\begin{theorem}[Over-exposed Users]\label{thm:over-exposure}
If $n_j$ denotes the number of impressions of the ad campaign from user group $j$, then the expected count of over-exposed users (served more than $c$ times) is
\[
N - \sum_{i=1}^N F\left(c;\sum_{j \in G_i}n_j, \frac{1}{k}\right).
\]
\end{theorem}

\subsection{Relaxing the Uniformity Assumption}\label{sec:non_uniform}
The prior sections rely on a key assumption that each user within a group has an equal probability of generating a group request. While this simplification is essential for privacy protection, it doesn't accurately capture real-world scenarios. However, under specific conditions, specifically when user groups do not overlap and advertisers define their target audience at a level less granular than these groups, we can improve ad performance without compromising $k$-anonymity.

Let's consider a scenario in which the ad platform can estimate each individual user's visit frequency. In such a case, we could associate each group $j$ with a property vector $(q_{j1}, q_{j2}, \ldots, q_{jk})$, where $q_{j1} \leq q_{j2} \leq \ldots \leq q_{jk}$ and $\sum_l q_{jl} = 1$. This vector represents the user visit probabilities within group $j$. To maintain $k$-anonymity, the ad system is only permitted to access these probabilities without linking them to specific users. For instance, a group with a property vector of $(0, 0, . . . , 0, 1)$ would signal to the ad system that there is a single active user in the group, while the rest are inactive; however, the system would not be able to identify the active user. The following theorem provides the expected reach for a user group in this setup.

\begin{theorem}\label{thm:expected_reach_nonuniform}
Given $n_j$ as the number of impressions of the ad campaign from user group $j$ with property vector $(q_{j1}, q_{j2}, \ldots, q_{jk})$, the expected reach generated from user group $j$ under frequency cap $c$ can be expressed as:
\begin{equation*}
\sum_{i=1}^k \Bigg(\sum_{l=1}^{c} lf\left(l;n_j, q_{ji}\right) + c - cF\left(c;n_j, q_{ji}\right)\Bigg).
\end{equation*}
\end{theorem}

The total reach of the campaign can then be derived by summing up the reach from each group, as we are assuming the groups do not overlap. In the case of unique reach ($c = 1$), we can state the following:
\begin{corollary}
Given $n_{j}$ as the number of impressions of the ad campaign from user group $j$ with property vector ($q_{j1}$, $q_{j2}$, . . . , $q_{jk}$), the expected number of unique reach from user group $j$ can be defined as:
\begin{equation*}
k-\sum_{i=1}^{k}(1-q_{ji})^{n_{j}}.
\end{equation*}
\end{corollary}

\section{Reach Optimization}
\label{section:optimization}
Measurement and optimization play pivotal roles in online advertising platforms, as they are essential tools that help marketers achieve a solid return on investment. Reach measurement is typically displayed on an ad reporting page within the publisher platform, allowing marketers to identify gaps in audience coverage and evaluate the effectiveness of their marketing strategies. Meanwhile, reach optimization is often offered by the platform through an automated bidding service, which places bids on behalf of advertisers in each auction to optimize results. As the outcome is reported by the measurement component, it is crucial that these two elements maintain consistency with one another. 

In this section, we focus on the reach maximization problem with regard to an ad campaign, which comprises an ad, a budget, and a targeted user set. In traditional scenarios where user identity is readily available, the advertising platform tracks the frequency of ad impressions per user. It halts the campaign for a given user once the frequency cap $c$ is reached. Here, we expand the reach optimization problem to account for ad requests originating from user groups.

Let $T$ denote the total number of ad impressions initiated by the targeted users' viewing or search sessions. Each impression is generally auctioned either via a first-price or second-price auction, with the winning bid claiming the impression. The objective of an ad campaign is to secure as many impressions as possible while ensuring that each targeted user receives the ad at most $c$ times. For instance, when $c = 1$, the goal is to reach as many unique targeted users as possible.

Consider the $t$-th impression and define $W_t(b)$ as the winning probability given a bid price of $b$ and $H_t(b)$ as the expected cost. Let their respective derivatives be $w_t$ and $h_t$. Let $B$ denotes the campaign's budget. We solve the following optimization problem to optimize reach:
\begin{equation}\label{eq:general-formulation}
\max_{b_t} \sum_{t=1}^T p_tW_t(b_t),\quad s.t.\ \sum_{t=1}^T H_t(b_t) \le B,
\end{equation}
where $p_t$ is the likelihood that the $t$-th impression originates from a targeted user who has seen the campaign's ad fewer than $c$ times. In a deterministic environment, if the user generating the impression is known, $p_t$ is binary, thus recovering the traditional frequency capping mechanism. Section \ref{section:impression_valuation} explains how to calculate $p_t$ in a stochastic scenario where only the user's group is identifiable. The term $p_tW_t(b_t)$ specifies the marginal contribution of the $t$-th impression, so this natural probabilistic frequency capping extension directly relates to the campaign's optimization objective.

Let $\lambda$ be the Lagrangian multiplier for the budget constraint, then the optimal bid $b_t^*$ given $\lambda$ is
\begin{equation}\label{eq:b-star}
b_t^*(\lambda) = \left(\frac{h_t}{w_t}\right)^{-1}\left(\frac{p_t}{\lambda}\right),
\end{equation}
where $(\cdot)^{-1}$ signifies function inverse.
The optimal Lagrangian multiplier $\lambda^*$ satisfies the KKT condition
\[
\sum_{t=1}^T H_t(b_t^*(\lambda^*)) = B.
\]
In practical terms, $\lambda^*$ can be obtained via an online solver \citep{gao2022bidding}.

\subsection{Estimating Reach Probability}\label{section:impression_valuation}

As discussed in the previous section, the crucial quantity in Equation \ref{eq:b-star} is $p_t$. This is the probability that the request comes from a user who has viewed the ad fewer than $c$ times.

Consider a campaign focused on maximizing unique reach ($c = 1$), within a user group composed of two targeted users: A and B. Suppose the group generates two ad requests indexed by $1$ and $2$. For the first request, $p_1 = 1$ since it is the initial time we encounter this group. If the campaign secures this impression, the next time this group visits the publisher, we know that either A or B has seen the ad. Hence, there is only a 50\% chance that we can reach a unique user, and therefore, $p_2 = 0.5$.

Formally, let $n_{tj}$ denote the number of impressions of the ad campaign from user group $j$ up until the $t$-th ad request. Let $j_t$ represent the index of the user group that generates the $t$-th ad request. We then have the following formula for $p_t$.

\begin{theorem}[Probability of Additional Reach]\label{thm:reach_probability}
Assume that user group $j_t$ generates ad request $t$. The probability $p_t$ that ad request $t$ comes from a user who has viewed the ad fewer than $c$ times is given by
\begin{equation}\label{eq:reach_probability}
p_t = \frac{1}{k}\sum_{i \in U_{j_t}}F\left(c - 1; \sum_{j \in G_i}n_{tj}, \frac{1}{k}\right).
\end{equation}
\end{theorem}

As a remark, as $c \rightarrow \infty$, we find that $p_t \rightarrow 1$, in which case, the optimization problem in \eqref{eq:general-formulation} simplifies to an impression volume optimization. For the specific case where $c = 1$, we have the following corollary.

\begin{corollary}\label{cor:optimization_unique_reach}
For a campaign optimizing for unique reach, we have
\[
p_t = \frac{1}{k}\sum_{i \in U_{j_t}}\left(1 - \frac{1}{k}\right)^{\sum_{j \in G_i} n_{tj}}.
\]
\end{corollary}

It's noteworthy that the probability $p_t$ decreases exponentially with $n_{tj}$ for a specific group $j$. To see how the winning probability diminishes as the group impression count increases, consider $p_t = \exp(-\alpha n_{tj})$, where $\alpha > 0$ is a parameter that depends on the group size. At the extremes, for a group of size $k = 1$, we have $\alpha \rightarrow +\infty$, and $\alpha \rightarrow 0$ as the group size $k \rightarrow \infty$.

Assuming a winning probability model in which the bids of competing campaigns follow a log-normal distribution for the $t$-th request from group $j$, i.e., $W_t(b) = \Phi((\ln b - \mu_t) / \sigma_t)$, the relationship between the winning probability and the number of impressions for group $j$ can be expressed as $W_t(p_t/\lambda) = \Phi((-\alpha n_{tj} - \ln \lambda - \mu_t)/\sigma_t)$. Generally, when the groups overlap, impressions from other groups should also be taken into account, but the decay is of the same order.

Finally, $p_t$ can be computed in a streaming fashion with time complexity $O(kc)$ and space complexity $O(cT)$.

\subsection{Non-uniform Setting}
For completeness, we also present the reach probabilities $p_t$ in the non-uniform arrival setting described in Section \ref{sec:non_uniform}.

\begin{theorem}\label{thm:reach_probability_nonuniform}
Let $n_{tj}$ be the number of impressions of the ad campaign from user group $j$ with property vector $(q_{j1}, q_{j2}, \ldots, q_{jk})$, up until ad request $t$ happens. Now suppose user group $j$ generates the $t$-th request, then the probability $p_t$ that ad request $t$ is from a user that has been served less than $c$ times previously is
\begin{equation}\label{eq:reach_probability_nonuniform}
p_t = \sum_{i=1}^k q_{ji}F\left(c - 1; n_{tj},  q_{ji}\right).
\end{equation}
\end{theorem}

In particular, for unique reach, $p_t = \sum_i q_{ji}(1 - q_{ji})^{n_{tj}}$.

\section{Experiments}
\label{section:experiments}

This section outlines the experiments conducted to address a range of significant questions. Firstly, we deploy Monte Carlo simulations on the reach distribution, demonstrating the accuracy of the reach estimate and its variance progression throughout a campaign's lifespan. Secondly, we show the increase in measurement uncertainty as stricter user privacy is implemented. Next, we run tests on campaign reach optimization under varying privacy levels, resulting in an efficiency versus privacy tradeoff curve. This quantifies the influence of privacy on digital brand advertising performance. Lastly, we also compare the performance of different methods on real-world datasets where user arrival within a group is non-uniform.

\subsection{Monte Carlo Simulations on Reach}

\begin{figure*}[t]
\centering
\includegraphics[trim=0 275 0 0, clip, width=0.95\linewidth]{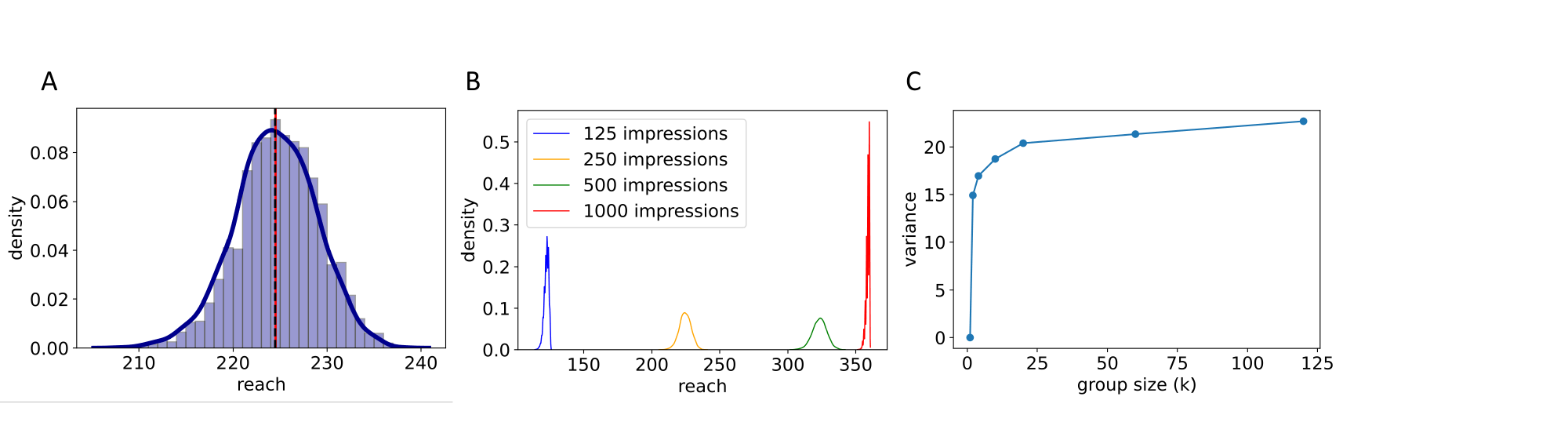}
\caption{Experiments on reach measurement. Reach is defined as the number of impressions under the frequency cap. A: Distribution of reach with 250 impressions and a frequency cap of 3. The black dashed vertical line indicates the mean of the sampled distribution. The red solid line depicts the expected reach calculated from Theorem \ref{thm:measurement}. B: Distributions of reach under different numbers of impressions. C: Variance of reach with different group sizes.}
\label{fig1}
\end{figure*}

The expected reach for a given set of group impressions is detailed in Theorem \ref{thm:measurement}. In this initial experiment, Monte Carlo simulations are conducted to derive the complete reach distribution. We model a campaign with 120 targeted users and a frequency cap, $c = 3$. These targeted users are divided into 20 groups, each containing $k = 6$ members. We randomly draw 250 impressions from these groups and sample the potential user trajectories that generated these impressions. This data is used to plot the reach distribution (Figure \ref{fig1}A), with the empirical mean from this distribution closely aligning with the expected reach from Theorem \ref{thm:measurement}.

Further analysis of the reach distribution evolution as impressions increase is illustrated in Figure \ref{fig1}B. When impressions are scarce, it is likely that most fall under the frequency cap and are counted as reach, resulting in a small distribution variance. As impressions accrue, the variance initially increases, but decreases as impressions become plentiful. This is due to the reach becoming saturated as a result of the frequency cap amidst a high number of impressions.

Lastly, the variances across various group sizes are plotted in Figure \ref{fig1}C. This illustrates the effect of heightened user privacy on the reach measurement's uncertainty. At $k = 1$, the measurement is deterministic. As user privacy is introduced ($k = 2$), a substantial increase in variance occurs. However, as we provide more user privacy, the growth in measurement uncertainty is minimal.

\subsection{Privacy vs. Efficiency Tradeoff}

Digital privacy is ushering in substantial changes within the advertising industry. There is a prevailing consensus that the efficacy of online marketing risks being compromised as stronger privacy protection is enacted. In this section we quantify the decline in brand advertising performance under different privacy levels.

In this experiment, we construct a campaign that targets 120 users who generate a total of 2000 ad impressions. These impressions are sold via second-price auctions hosted by the advertising platform. The winning price (second price) of each request is drawn from a log-normal distribution with $\mu = 0$ and $\sigma^2 = 0.5$. The advertising platform maps these targeted users into groups of size $k$, and only the group id is made available for each ad request. A bidding agent places bids on behalf of the advertiser, adhering to the bidding formula \eqref{eq:b-star}. The $\lambda_t$ is then updated based on the following online algorithm, as suggested in \citep{balseiro2019learning, gao2022bidding}:
\[
\lambda_{t+1} = \lambda_{t}-\eps(1-r_t),
\]
where $r_t$ signifies the ratio of the spend in request $t$ and the average budget per request $B/T$. The learning rate $\eps$ is set to $0.1$. To imitate real-world scenarios in online production systems, lower and upper bounds on bid prices are also established at $0.1$ and $10$, respectively. The starting bid $\lambda_1$ is fixed at $10$.

\begin{figure}[t]
\centering
\includegraphics[trim=0 250 475 0, clip, width=0.95\linewidth]{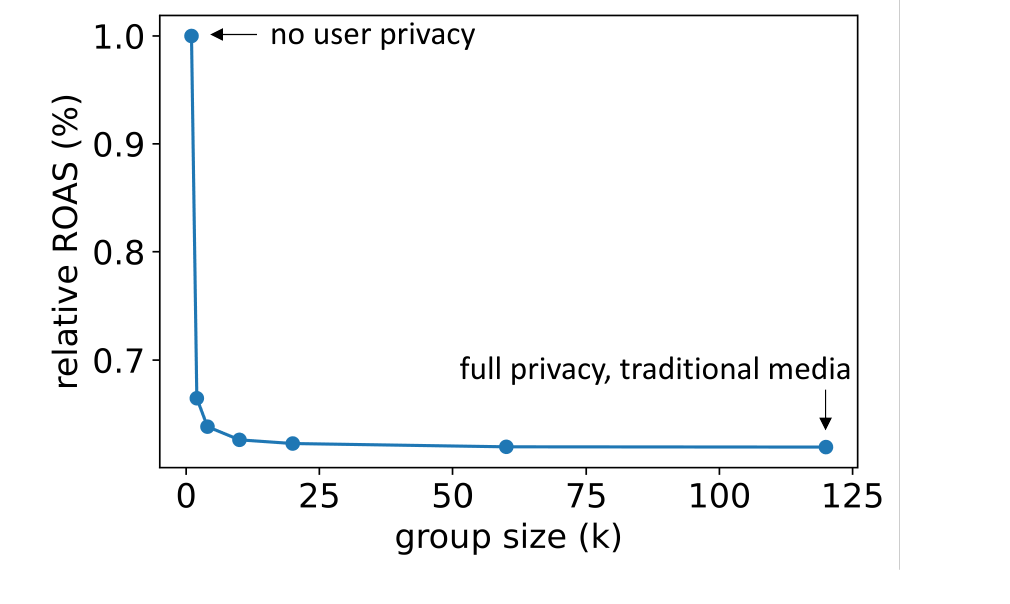}
\caption{Privacy vs. efficiency in online brand advertising. The relative ROAS is plotted against different group sizes $k$ representing different levels of user privacy.}
\label{fig2}
\end{figure}

The return on ad spend (ROAS), here defined as the ratio of total reach over total spend, is plotted against varying group sizes (Figure \ref{fig2}). When no user privacy is in place ($k = 1$), the best possible ROAS is achieved. ROAS for other scenarios is normalized against this optimal upper limit, facilitating easier comprehension of the performance decline percentage. As anticipated, performance deteriorates monotonically as $k$ increases. Notably, a 33\% drop in ROAS is observed as user privacy is introduced (shifting from $k = 1$ to $k = 2$). However, there isn't a significant additional cost for enhanced privacy. In the extreme case when all users form a single group, the situation mirrors advertising on traditional media, where frequency capping is only applicable to the entire audience population. Here, the efficiency is merely 62\% of fully personalized online advertising.

\subsection{Evaluation on Real-world Dataset}
\label{sec:realworld}

\begin{figure}[t]
\centering
\includegraphics[trim=0 175 175 0, clip, width=0.95\linewidth]{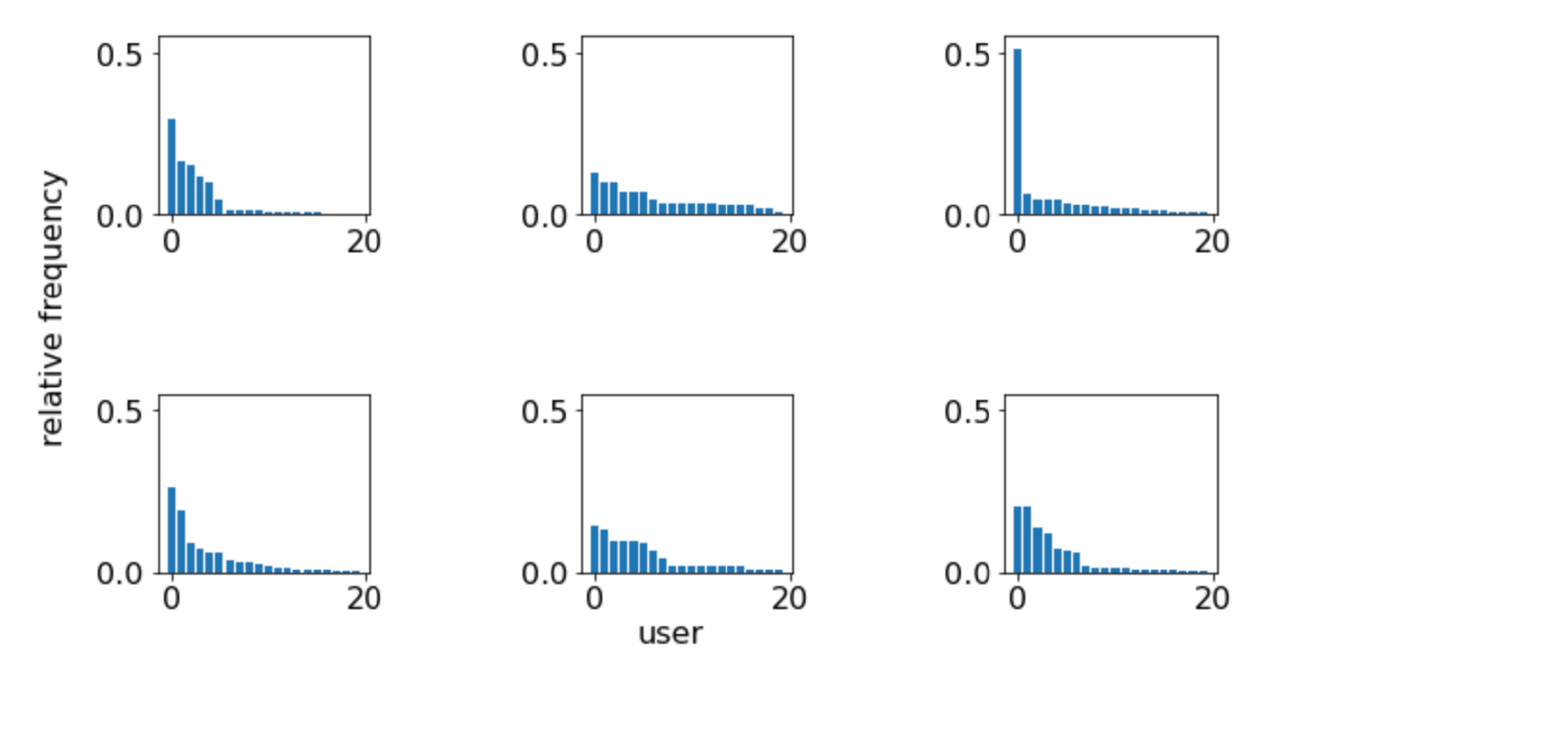}
\caption{Relative frequency of each user within a group. Distributions of six groups are shown. Users are ranked based on their frequency.}
\label{fig5}
\end{figure}

In practice the number of user impressions within a group can vary a lot. To bridge this gap, further experiments were conducted using auction and impression data from a real-world advertising platform. Users on the platform were randomly assigned to non-overlapping groups of size 20. Frequency counts of each member within a group revealed a variation in their distribution across different groups, as demonstrated in Figure~\ref{fig5}.

In this experiment, we focus specifically on unique reach ($c = 1$). We evaluated four unique reach measurement and optimization approaches on a total of 10 ad campaigns. These approaches include:

\begin{itemize}
\item \textbf{A. Unique impressions:} Each impression from a user group is considered a reach, leading to an overestimation.
\item \textbf{B. Unique groups:} Reach is defined as the number of unique group impressions, resulting in an underestimation.
\item \textbf{C. Uniform reach:} Reach is estimated using Theorem 1 and reach probability is calculated via Theorem 4, which assumes a uniform distribution of user frequencies within a group.
\item \textbf{D. Non-uniform reach:} Reach is estimated based on Theorem 5, and reach probability is calculated via Theorem 6. For each group, the property vector is derived from actual user visit frequencies in the dataset.
\end{itemize}

\begin{figure}[t]
\centering
\includegraphics[width=\linewidth]{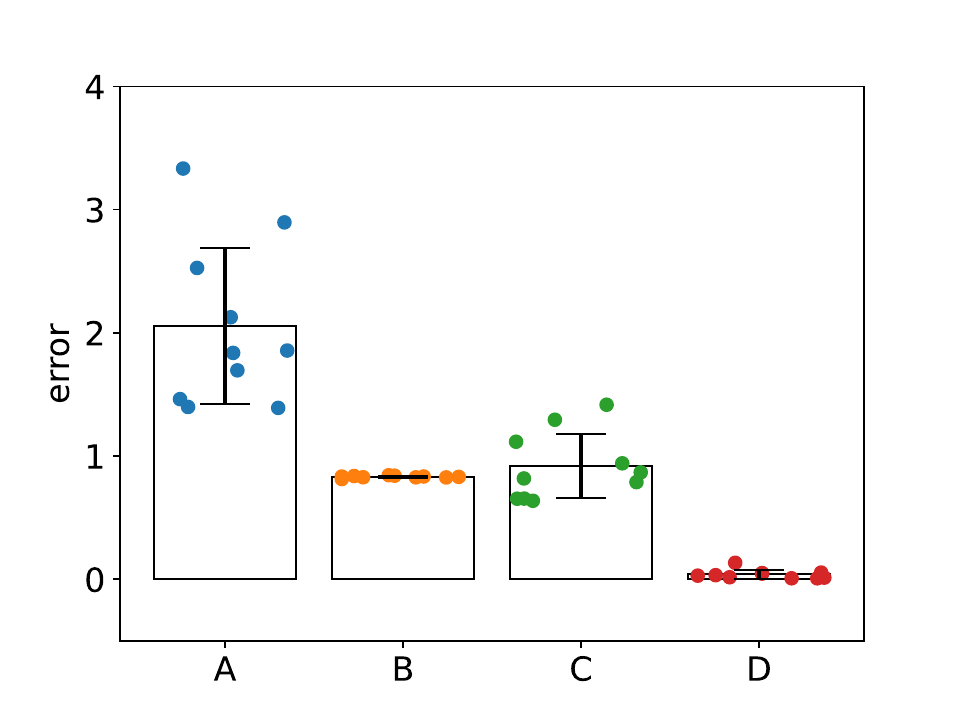}
\caption{Reach measurement performance comparison. The error metric of each campaign is plotted as a dot. Error bars represent standard deviation. A, B, C, D represents the four approaches listed in section \ref{sec:realworld}.}
\label{fig6}
\end{figure}

Using $R_{\text{true}}$ as the ground truth reach and $R$ as the measured reach, we calculated the relative error of reach measurement for all methods using the formula $|R - R_{\text{true}}| / R_{\text{true}}$. As illustrated in Figure~\ref{fig6}, approach D provides the most accurate measurement among the four methods. It's worth noting that for D, we did not account for estimation errors in the property vectors. However, these can usually be forecasted with a reasonable level of accuracy using historical user visit data. Importantly, approach C (which assumes a uniform distribution within the group) consistently overestimates the actual reach. This is because maximizing (9) in the probability simplex reaches its maximum with uniform distribution.

To evaluate the reach optimization performance of these four approaches, we used the same bidding algorithm described in Section 5.2, discounting the bids at each request level using the reach probabilities calculated from the different approaches. For each campaign, its bids in the auctions were replaced by the output of the bidding algorithm, while other competing bids remained unchanged. For all methods, we measured the relative ROAS compared to existing performance where no privacy is enforced. The results are summarized in Table \ref{tab:results}.

\begin{table}[t]
\centering
\begin{tabular}{|c|c|c|c|c|}
\hline
& A & B & C & D \\
\hline
Mean & 85.99\% & 88.42\% & 86.95\% & 94.12\% \\
\hline
Variance & 0.023 & 0.018 & 0.023 & 0.005 \\
\hline
\end{tabular}
\caption{Mean and variance of relative ROAS of the different approaches across the 10 campaigns. A, B, C, D represents the four approaches listed in Section \ref{sec:realworld}.}
\label{tab:results}
\end{table}

It's evident that relaxing the uniformity assumption within user groups results in improved measurement accuracy and advertising performance, primarily because this model better corresponds with real-world user visit patterns. However, this approach necessitates that user groups do not overlap and all calculations are conducted at the group level.

\section{Final Remarks}

To the best of our knowledge, this paper is the first to investigate reach measurement, optimization, and frequency capping under the $k$-anonymous privacy protocol. We outline how to report reach in this new privacy context and demonstrate how probabilistic discounting, a probabilistic variant of traditional frequency capping, can be leveraged to optimize campaign performance. For ease of presentation, we have assumed that all user groups are of equal size $k$. However, the findings are adaptable to scenarios where user groups vary in size.

The privacy landscape is fast-evolving and it has profound implications on the future of digital advertising. This study evaluates one of the many possible privacy protocols. Comparing the advertising efficiency vs. user privacy tradeoffs of different privacy frameworks is an important future research direction.

Finally, the comparative analysis of various methodologies for optimizing reach in real-world datasets underscores the enhanced performance achieved by relaxing the uniform assumption. This finding points to an encouraging path forward: incorporating realistic models of user behavior into the design of reach optimization strategies, especially within the context of $k$-anonymous environments, presents a promising avenue for advancing the efficacy and precision of targeted online advertising while upholding user privacy.

\bibliographystyle{ACM-Reference-Format}
\bibliography{neurips_2024}

\appendix

\section{Supplement Materials}

\subsection{Proofs}

\subsubsection{Proof of Theorem \ref{thm:measurement}}

Let $X_i$ be the random variable for the number of times user $i$ has seen the ad. For any impression generated from a user group that contains user $i$, there is a $1/k$ probability that the ad request is from user $i$. This corresponds to a Bernoulli trial with a success probability $1/k$. Therefore within the given time window, $X_i \sim \mathcal B\left(\sum_{j \in G_i} n_j, 1/k\right)$, a Binomial distribution with the number of trials as the sum of impressions generated from any user group that contains user $i$.

Denote $R(c)$ as the random variable for the number of impressions under frequency cap $c$, and $R_i(c) := \min(X_i, c)$ the number of impressions under frequency cap $c$ for user $i$. Then $\E[R(c)] = \sum_i \E[R_i(c)] := \sum_i \E[\min(X_i, c)]$. The result is then obtained by direct calculation, i.e.,
\begin{align*}
\E[R(c)] &= \sum_{i=1}^N\E[\min(X_i, c)] \\
&= \sum_{i=1}^N \E[X_i | X_i \le c] + cP(X_i > c)\\
&= \sum_{i=1}^N \Bigg(\sum_{l=1}^{c} lf\left(l;\sum_{j \in G_i}n_j, \frac{1}{k}\right) + c - cF\left(c;\sum_{j \in G_i}n_j, \frac{1}{k}\right)\Bigg).
\end{align*}

\subsubsection{Proof of Corollary \ref{cor:measurement_unique_reach}}

This naturally comes as a special case of Theorem \ref{thm:measurement}, but here we provide an alternative proof that is more intuitive. Define $Y_i = \min(X_i, 1)$, then $E[Y_i] = P(Y_i = 1)$, the probability that user $i$ is reached at least once. With this probabilistic interpretation the conclusion is obvious.

\subsubsection{Proof of Theorem \ref{thm:concentration}}

We first prove Lemma \ref{lem:cov}. The expectation of $R_i$ is given in \eqref{eq:unique_reach_measurement}, i.e.,
\[
E[R_i] = \left(1 - \left(1 - \frac{1}{k}\right)^{\sum_{j \in G_i}n_j}\right).
\]
With inclusion–exclusion principle, we have
\begin{equation*}
\begin{split}
P(R_iR_{i'} = 0) &= P(R_i=0) + P(R_{i'}=0) - P(R_i=0, R_{i'} = 0)\\
&= \left(1 - \frac{1}{k}\right)^{\sum_{j \in G_i}n_j} + \left(1 - \frac{1}{k}\right)^{\sum_{j \in G_{i'}}n_j} \\
&\quad -\left(1 - \frac{1}{k}\right)^{\sum_{j \in G_\cup - G_\cap}n_j} \left(1 - \frac{2}{k}\right)^{\sum_{j \in G_\cap}n_j}
\end{split}
\end{equation*}
Note that $\E[R_iR_{i'}] = P(R_iR_{i'} = 1) = 1 - P(R_iR_{i'} = 0)$,
Therefore
\begin{equation*}
\begin{split}
\Cov[R_i, R_{i'}] &= \E[R_iR_{i'}] - \E[R_i]\E[R_{i'}]\\
&= \left(1 - \frac{1}{k}\right)^{\sum_{j \in G_\cup - G_\cap}n_j} \\
&\quad \times \left(\left(1 - \frac{2}{k}\right)^{\sum_{j \in G_\cap}n_j}- \left(1 - \frac{1}{k}\right)^{2\sum_{j \in G_\cap}n_j}\right).
\end{split}
\end{equation*}
This concludes the proof of Lemma \ref{lem:cov}. Finally, the covariance formula allows us to calculate the variance of reach, i.e.,
\[
\Var[\sum_i R_i] = \sum_i \Var[R_i] + \sum_{i \neq i'}\Cov[R_i, R_{i'}],
\]
and Theorem \ref{thm:concentration} is derived by applying Chebyshev's inequality.

\subsubsection{Proof of Theorem \ref{thm:over-exposure}}
The number of users who are over-exposed is simply $\sum_{i=1}^N P(X_i > c)$, so the result is straightforward given the distribution of $X_i$.
\subsubsection{Proof of Theorem \ref{thm:expected_reach_nonuniform}}
Let $Z_{i}$ represent the random variable for the number of times the user with visit probability $q_{ji}$ has viewed the ad. In this case, $Z_{i} \sim \mathcal B(n, p)$. The proof follows a similar line of reasoning to that of Theorem \ref{thm:measurement}.
\subsubsection{Proof of Theorem \ref{thm:reach_probability}}
Let $X_i$ denote the random variable for the number of times user $i$ has seen this ad. Then $p_t$, the probability that ad request $t$ originates from a user who has viewed the ad fewer than $c$ times, is simply
\[
p_t = \frac{1}{k}\sum_{i \in U_{j_t}} P(X_i < c),
\]
which is an average of the probabilities for each user in group $j_t$. As shown in the proof of Theorem \ref{thm:measurement}, $X_i$ follows a Binomial distribution with parameters $\sum_{j \in G_i} n_{tj}$ and $1/k$, leading us to \eqref{eq:reach_probability}.
\subsubsection{Proof of Theorem \ref{thm:reach_probability_nonuniform}}
Let $Z_{i}$ represent the random variable for the number of times the user with visit probability $q_{ji}$ has viewed the ad. In this case, $Z_{i} \sim \mathcal B(n, p)$. The proof then follows a similar line of reasoning to that of Theorem \ref{thm:reach_probability}.

\subsection{Algorithms}

\subsubsection{Streaming Algorithm for Reach Measurement}

Theorem \ref{thm:measurement} allows the publisher to compute reach metrics given any time window. We give a streaming algorithm (Algorithm \ref{alg:measurement}) that outputs the expected number of reach as new impressions are getting observed in real time. The algorithm is based on Corollary \ref{cor:alternative_form}.

\begin{algorithm}[tb]
\caption{Streaming Algorithm For Reach Measurement}
\label{alg:measurement}
Set $n[i] = 0$ for $i = 1, \ldots, N$\;
Set $f[l, t] = 0,$ for $l = 0, \ldots, c; t = 0, \ldots, T$\;
Set $f[l, l] = \left(\frac{1}{k}\right)^l$ for $l = 0, \ldots, c$\;
Set $R[m]= 0,$ for $m = 1, \ldots, c$\;
\For{$t = 1, \ldots, T$}{
    \If{the campaign wins request $t$ from group $j_t$}{
        \For{$i$ in $U_{j_t}$}{
            Increment $n[i]$ by $1$\;
            \For{$l = 0, \ldots, c$}{
                \If{$n[i] > l$}{
                    \If{$f[l, n[i]] = 0$}{
                        Set $f[l, n[i]] = \frac{f[l, n[i] - 1]n[i]}{n[i] - l}\left(1 - \frac{1}{k}\right)$\;
                    }
                }
                \For{$m = l+1, \ldots, c$}{
                    Increment $R[m]$ by $(m-l)(f[l,n[i]-1]-f[l,n[i]])$\;
                }
            }
        } 
    }
    \For{$m = 1, \ldots, c$}{
        Output $R[m]$, the reach at frequency cap $m$\;
    }
}
\end{algorithm}

The time complexity of Algorithm \ref{alg:measurement} for each new impression is $O(kc^{2})$, and the overall space complexity is $O(cT)$. It uses dynamic programming to avoid duplicate computation of $f(l; n, 1/k), \forall l \in \{1, \ldots, c\}, n \in \{1, \ldots, T\}$. For the particular case when the groups are non-overlapping, the time complexity goes down to $O(c^2)$ as the computation can be conducted on the group level instead.

\subsubsection{Streaming Algorithm for Reach Optimization}

Similar to the measurement algorithm, a streaming algorithm that computes $p_t$ is given in Algorithm \ref{alg:optimization}. The quantity $p_t$ shows up in the bidding formula \ref{eq:b-star}, which is used to optimize campaign performance.

\begin{algorithm}[tb]
\caption{Streaming Algorithm for reach probability $p_t$}
\label{alg:optimization}
Set $n[i] = 0$ for $i = 1, \ldots, N$\;
Set $f[l, t] = 0,$ for $l = 0, \ldots, c; t = 0, \ldots, T$\;
Set $f[l, l] = \left(\frac{1}{k}\right)^l$ for $l = 0, \ldots, c$\;
\For{$t = 1, \ldots, T$}{
    Set $p_t = 0$\;
    \For{$i$ in $U_{j_t}$}{
        Increment $p_t$ by $\frac{1}{k}\sum_{m=0}^{c-1}f[m, n[i]]$\;
    }
    Output $p_t$\;
    \If{the campaign wins request $t$ from group $j_t$}{
        \For{$i$ in $U_{j_t}$}{
            Increment $n[i]$ by $1$\;
            \For{$l = 0, \ldots, c$}{
                \If{$n[i] > l$}{
                    \If{$f[l, n[i]] = 0$}{
                        Set $f[l, n[i]] = \frac{f[l, n[i] - 1]n[i]}{n[i] - l}\left(1 - \frac{1}{k}\right)$\;
                    }
                }
            }
        }
    }
}
\end{algorithm}

The time complexity for each request is $O(kc)$, and the overall space complexity is $O(cT)$.

\section{Additional Experiment on the Impact of Targeted User Coverage}

In our experiments, we operated under the assumption that all users within a group are targeted. However, in actual advertising scenarios, advertisers target users based on various attributes such as geolocation, age, and gender. The alignment of these targeting attributes to a set of groups may not always be perfect, depending on how the user group mapping is structured. This can lead to situations where only a subset of users within a group is targeted.

In this additional experiment, we evaluate the influence of the extent of targeted user coverage within groups. We setup a campaign that targets only 12 users within a population of 144 users. These 144 users are divided into 12 user groups. We manipulate how the 12 targeted users are distributed within these 12 groups. In one extreme, all of them are housed within a single group. Conversely, in the other extreme, each of them belongs to a unique user group. We define coverage as the average proportion of targeted users within a group, considering only groups that contain at least one targeted user. In the extreme cases, the coverage levels are 100\% and 8.33\%, respectively.

\begin{figure}[t]
\centering
\includegraphics[width=\linewidth]{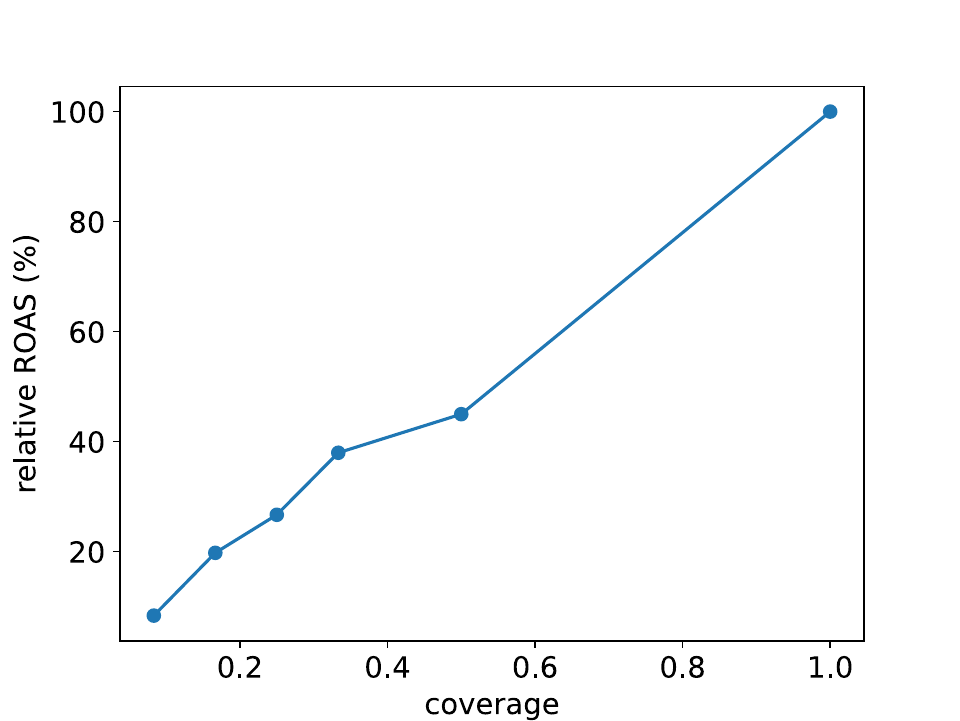}
\caption{Impact of the coverage of targeted users in groups. Coverage is defined as the average percentage of the targeted users in a group. A coverage of $1.0$ implies that all the users in the group are within the target.}
\label{fig7}
\end{figure}

A plot of ROAS (return on ad spend) versus coverage demonstrates an almost linear increase in efficiency with coverage, underlining the significance of creating groups that align with advertisers' targeting segments (Figure~\ref{fig7}).

Combining the observation from other experiments, we emphasize that the erosion of ROAS under $k$-anonymity is attributed to several factors. Firstly, the enforcement of frequency capping in user groups now can only be probabilistic, leading to potential wastage of advertising budget on over-exposed users. Secondly, the reach probability $p_t$ for a user group can be significantly reduced by a few active users within the group, preventing the campaign from reaching the other users in the group. Lastly, when a group contains a mix of targeted and untargeted users, there's a risk of ads being served to users outside the target range, resulting in further wastage of advertising spend.

\end{document}